\documentclass[aps,prl,twocolumn,amsmath,amssymb]{revtex4}

\usepackage{graphicx}
\usepackage{dcolumn}
\usepackage{bm}
\usepackage{subfigure}
\usepackage{multirow}

\begin{document}

\title{Stochastic robustness and relative stability of multiple pathways in biological networks}

\author{Yongyi Guo$^1$}
\author{Zhiyi You$^1$}
\author{Min Qian$^1$}
\author{Hao Ge$^2$}
\email{haoge@pku.edu.cn}
\affiliation{
$^1$School of Mathematical
Sciences, Peking University, Beijing, 100871, PRC.\\
$^2$Beijing International Center for Mathematical Research and Biodynamic \\Optical Imaging Center, Peking University, Beijing, 100871, PRC.}

\date{\today}

\begin{abstract}
Multiple dynamic pathways always exist in biological networks, but their robustness against internal fluctuations and relative stability have not been well recognized and carefully analyzed yet. Here we try to address these issues through an illustrative example, namely the Siah-1/beta-catenin/p14/19 ARF loop of protein p53 dynamics. Its deterministic Boolean network model predicts that two parallel pathways with comparable magnitudes of attractive basins should exist after the protein p53 is activated when a cell becomes harmfully disturbed. Once the low but non-neglectable intrinsic fluctuations are incorporated into the model, we show that a phase transition phenomenon is emerged: in one parameter region the probability weights of the normal pathway, reported in experimental literature, are comparable with the other pathway which is seemingly abnormal with the unknown functions; whereas, in some other parameter regions, the probability weight of the abnormal pathway can even dominate and become globally attractive. The theory of exponentially perturbed Markov chains is applied and further generalized in order to quantitatively explain such a phase transition phenomenon, in which the nonequilibrium ``activation energy barriers" along each transiting trajectory between the parallel pathways and the number of ``optimal transition paths" play a central part. Our theory can also determine how the transition time and the number of optimal transition paths between the parallel pathways depend on each interaction's strength, and help to identify those possibly more crucial interactions in the biological network.


\end{abstract}


\maketitle

Biological entities have to function robustly against internal stochastic fluctuations for the
sake of survival and evolution \cite{LHTW,Stelling2014}. At the system level, robustness is a complex and emergent property, which could not be simply inferred through only examining the characteristics of each isolated component.

Hence, mathematical models have always been used to investigate the robustness of biological networks. Local stability can be predicted by the deterministic model: each stable state of the biological networks corresponds to an attractor in the state space, which means trajectories of the system originating from certain initial conditions nearby move towards it as time increases. However, it is common for a deterministic dynamical system to have multiple attractors, each of which has its own basin of attraction, defined as the set of initial states approaching that attractor asymptotically. Therefore, how to quantify the relative stability between different attractors becomes an important issue.

Internal stochastic fluctuations are unavoidable in biological networks. After stochasticity is incorporated, the relative stability can be quantitatively addressed according to the comparison of their probabilities, rather than attributing to the magnitudes of their own basins of attraction \cite{Zeeman1988}.

There are mainly two types of attractors: fixed points and limit cycles, which correspond to steady states and oscillatory pathways respectively in biological networks. The relative stability of multiple steady states have been carefully studied in many different stochastic models recently, and a nonequilibrium landscape theory has been proposed  \cite{Landscape}. However, the relative stability and stochastic robustness of parallel pathways have not been comparably recognized and analyzed, which indeed has already caused much interests from biologists \cite{Huang2006}.

\begin{figure}[bht]
\centerline{\includegraphics[width=7cm]{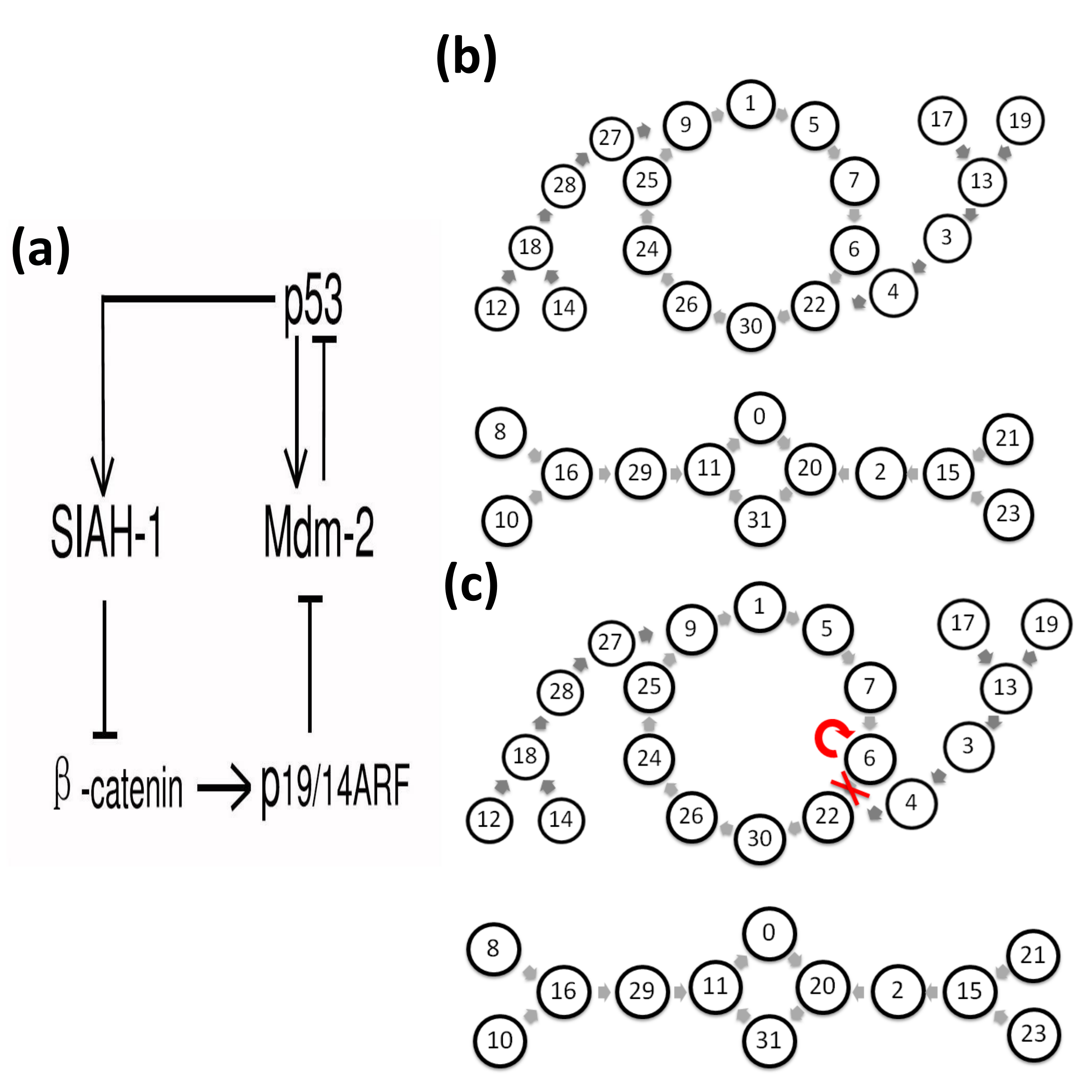}}
\caption{Siah-1/beta-catenin/p14/19 ARF loop and the dynamics of deterministic Boolean network model. (a) Siah-1/beta-catenin/p14/19 ARF loop, redrawn from \cite{HL}. (b,c) Visualization of the deterministic boolean network model with $\delta=1$ (b) and $\delta=0$ (c). The numbers 0-31 correspond to the 32 states by means of replacing ``0" with ``-1" in their respective binary representation.}
\label{fig0}
\end{figure}

In the present letter, we take the Siah-1/beta-catenin/p14/19 ARF loop of protein p53 dynamics as an illustrative example (Fig. \ref{fig0})\cite{HL}. The p53 protein after activated can positively regulate the transcription of the ubiquitin ligase Siah-1, which turns to degrade beta-catenin protein. On the other hand, the increasing of the beta-catenin level can activate the p14/19 ARF gene, which in turn negatively control MDM-2, resulting in further activation of p53 protein. We build both deterministic and stochastic Boolean network models for such a typical biological network, in which two parallel pathways emerge after the cell is damaged. We shall show that in different parameter regions, the relative stability of the parallel pathways behaves very differently, analogous to phase transition in statistical mechanics. These phenomenon can be well explained by a generalization of the theories of exponentially perturbed Markov chains.

Here, we apply the approach of Boolean network model, which has already been used to explore general and global properties of large gene expression networks \cite{Kauff91,Kauff93,AMK99,LLL04,ZQ1,GQQ_MBS_2008,H82,H84,Am}. The Boolean description of cellular states is due to the on-off characteristics of the components in the biological networks. The advantage of such a qualitative model is that the predicted results are often not very sensitive to unknown kinetic parameters \cite{DMMO}.

Recently, the stability of several protein p53
pathways has been investigated through stochastic Boolean network models \cite{GQ_JCB_2008}, in which the unique pathway predicted in the deterministic model is globally attractive and has a dominant cycle flux in the stochastic model, which is the natural generalization of the deterministic limit cycle. Here we apply the same Boolean network model to the biological diagram in Fig. \ref{fig0}(a), where the sequence of the nodes is (p53, Siah-1, $\beta$-catenin, p14/19 ARF, Mdm-2). In such a Boolean network model, each involved key factor (e.g. proteins, DNAs or RNAs) in the biological network of interest only has two states, $X_i=1$ or $X_i=-1$, representing its active and inactive states respectively. The states in the next time step are determined by the present states via the following rules \cite{LLL04,H82,H84,GQ_JCB_2008,GQQ_MBS_2008}:

Denote a matrix $T=\{T_{ij}\}$, in which $T_{ij}=1$ (activation) for an arrow from protein $j$ to protein $i$ in Fig. 1 (a), and $T_{ij}=-1$ (inhibition) for a horizontal bar instead of
arrowhead from protein $j$ to protein $i$ in Fig. 1 (a). Other elements are set as zero. Given $T$, the deterministic model can be described as follows: if $X(t)\neq(-1,-1,1,1,-1)$, then
\begin{eqnarray}
    X_i(t+1)=\left\{\begin{array}{ll}\textrm{sign}
    (H_i),&\textrm{if}~H_i\neq 0;\\X_i(t),&\textrm{if}~H_i=0;\end{array}\right.
\end{eqnarray}
where the function
$\textrm{sign}(x)=\left\{\begin{array}{ll}1&x>0\\0&x=0\\-1&x<0\end{array}\right.$, and
$H_i=\sum_{j} T_{ij}X_j(t)$ is the input to the $i$-th node.

Another parameter $\delta$ indicates whether the cell is under normal
environment ($\delta=0$) or suffered from certain external perturbation (signal) ($\delta=1$), e.g. DNA damage. According to the diagram in Fig. 1 (a), inside a normal cell, the level of p53 protein is quite low, which in turn results in a low expression of Siah-1; the inhibition of Siah-1 then activates the $\beta$-catenin as well as p19/14ARF, leading to the suppression of Mdm-2. Hence, the state $(-1,-1,1,1,-1)$ should be a fixed point of the deterministic dynamics with $\delta=0$, i.e. $X(t+1)=(-1,-1,1,1,-1)$ if $X(t)=(-1,-1,1,1,-1)$. Once the cells encounter certain external damage signal as indicated by $\delta=1$, the protein p53 is activated first, i.e. $X(t+1)=(1,-1,1,1,-1)$ if $X(t)=(-1,-1,1,1,-1)$.


\begin{table}[h]
\begin{center}
\begin{tabular}{ccccccc}
Time&p53&Siah-1&$\beta$-catenin&p14/19
ARF&Mdm-2&Node no.\\
\hline
1&-1&-1&-1&-1&1&1\\
2&-1&-1&1&-1&1&5\\
3&-1&-1&1&1&1&7\\
4&-1&-1&1&1&-1&6\\
5&1&-1&1&1&-1&22\\
6&1&1&1&1&-1&30\\
7&1&1&-1&1&-1&26\\
8&1&1&-1&-1&-1&24\\
9&1&1&-1&-1&1&25\\
10&-1&1&-1&-1&1&9
\end{tabular}
\end{center}
\caption[tab1]{Cycle 1 evolution of the states} \label{tab1}
\end{table}

\begin{table}[h]
\begin{center}
\begin{tabular}{ccccccc}
Time&p53&Siah-1&$\beta$-catenin&p14/19
ARF&Mdm-2&Node no.\\
\hline
1&-1&1&-1&1&1&11\\
2&-1&-1&-1&-1&-1&0\\
3&1&-1&1&-1&-1&20\\
4&1&1&1&1&1&31
\end{tabular}
\end{center}
\caption[tab2]{Cycle 2 evolution of the states} \label{tab2}
\end{table}

There are always two parallel attractors in the deterministic models, no matter $\delta=0$ or $1$ (Fig. \ref{fig0} (b) and (c) with node numbers 0-31 corresponding to the 32 states here by means of replacing ``0" with ``-1" in their respective binary representation). When $\delta=1$, the two attractors are both limit cycles (see tables \ref{tab1} and \ref{tab2} for cycle 1 and 2 respectively). The magnitude of the two cycles' attractive basins is comparable to each other (12 and 20 nodes respectively), preventing us from determining their relative stability at this stage. When $\delta=0$, Cycle 1 in the deterministic model is replaced by the fixed point $(-1,-1,1,1,-1)$ (Node number 6).


However, in fact only Cycle 1 matches the reported Siah-1/beta-catenin/p14/19 ARF loop of the p53 pathways in the biological literature \cite{HL}. Therefore, we must appeal to the stochastic approach if we would like to understand the relative stability of the parallel attractors and the transition time between them.

In the stochastic approach, the deterministic dynamics are replaced by the transition probability introducing temperature-like parameters $\alpha$, $\beta$ and $\gamma$ (modified from \cite{Am,ZQ1,GQ_JCB_2008,GQQ_MBS_2008}):
\begin{equation}
    P(X(t+1)|X(t))=\prod_{i} P(X_i(t+1)|X(t)),\label{Eq_pd}
\end{equation}
where
$$P(X_i(t+1)|X(t))=\frac{\exp(\beta X_i(t+1)H_i)}
    {\exp(\beta H_i)+\exp(-\beta H_i)},$$
if $H_i\neq 0$, and $X(t)\neq(-1,-1,1,1,-1)$ or $i\geq 2$,
$$P(X_i(t+1)|X(t))=\left\{\begin{array}{ll}\frac{1}{1+e^{-\alpha}},&X_i(t+1)=X_i(t);\\
\frac{e^{-\alpha}}{1+e^{-\alpha}},&X_i(t+1)=-X_i(t),\end{array}\right.$$
if $H_i=0$, and $X(t)\neq(-1,-1,1,1,-1)$ or $i\geq 2$, and
$$P(X_1(t+1)|X(t))=\left\{\begin{array}{ll}\frac{e^{-\gamma}}{1+e^{-\gamma}},&X_1(t+1)=X_1(t),\\
\frac{1}{1+e^{-\gamma}},&X_1(t+1)=-X_1(t),\end{array}\right.$$ if
$X(t)=(-1,-1,1,1,-1)$. Here $\gamma\in \mathbb{R}\cup \{\pm \infty\}$.

\begin{figure}[bht]
\centerline{\includegraphics[width=8.5cm]{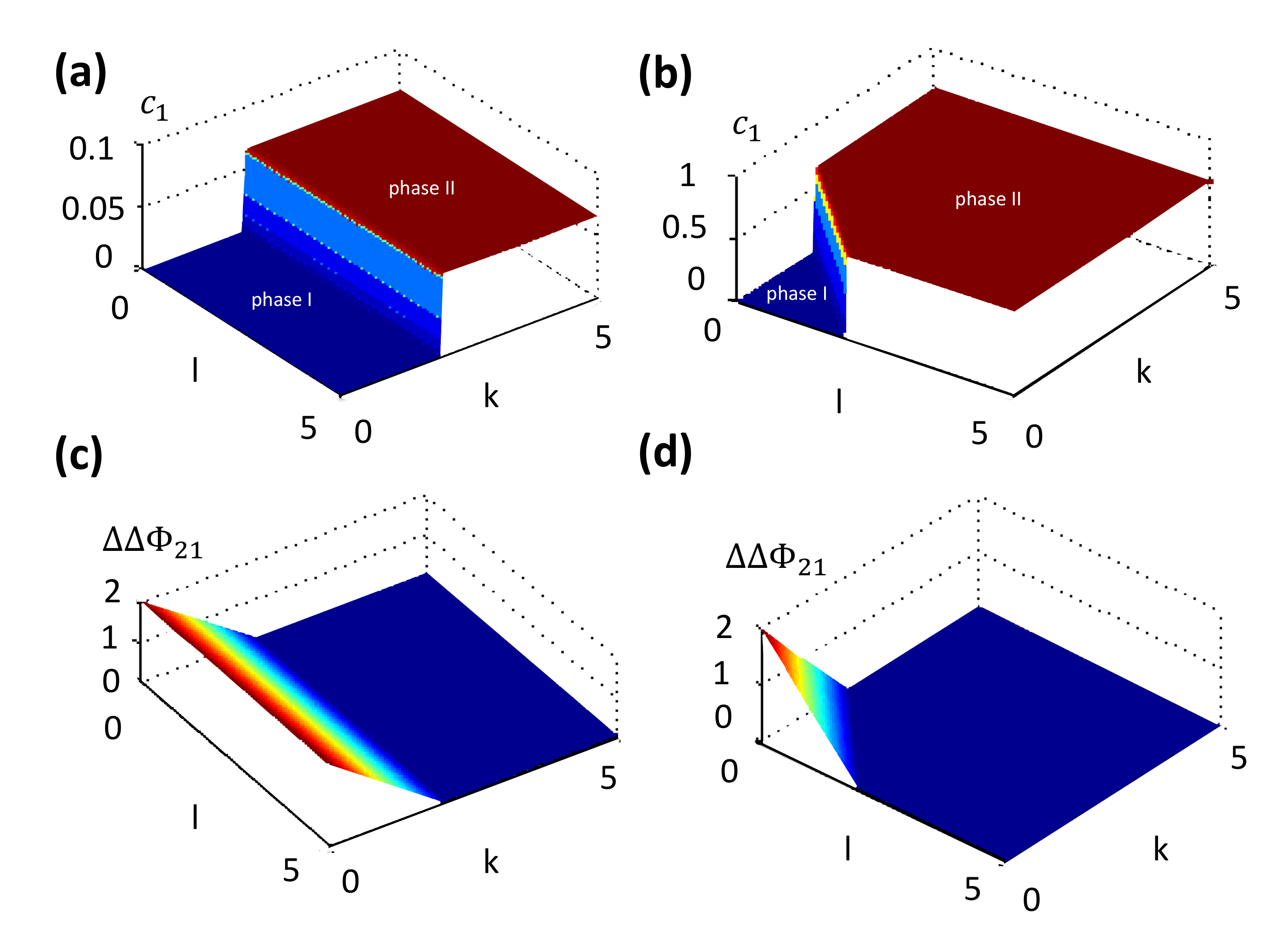}}
\caption{Limiting behavior of the stochastic model and its relation with the ``minimum activation energy barrier". (a) $\delta=1$: Net flux of Cycle 1 with respect to $k$ and $l$ given $\beta=50$. Here $\alpha=k\beta$ and $\gamma=l\beta$. (b) $\delta=0$: Net flux of Cycle 1 with respect to $k$ and $l$ given $\beta=50$. Here $\alpha=k\beta$ and $\gamma=-l\beta$. (c,d) Minimum activation energy barrier difference with respect to $k$ and $l$ corresponding to the cases of $\delta=1$ (c) and $\delta=0$ (d).}
\label{fig1}
\end{figure}

To better illustrate the behavior of the stochastic Boolean network model in the limit of vanishing fluctuation, we set in advance that $\alpha=k\beta$ and $\gamma=l\beta$ ($k$ and $l$ are nonnegative) when the stochastic model approaches the deterministic counterpart with $\delta=1$, as $\alpha$, $\beta$ and $\gamma$ are all very large. The cycle fluxes, which is the average times of occurrences of parallel pathways can be calculated accurately according to the expression in \cite{JQQ}. As long as $k>2$, the net flux of Cycle 1 is about $\frac{1}{15}$ regardless of the value of $l$ given a fixed and sufficiently large value of $\beta=50$ (Fig. \ref{fig1} (a)). Hence the probability of Cycle 1 is about $\frac{2}{3}$ that is two times more than that of the other cycle. But when $k<2$, the net flux of Cycle 1 nearly vanishes, while Cycle 2 becomes globally attractive with probability $1$, i.e. the forward transition time from Cycle $1$ to Cycle $2$ is exponentially small compared to the backward one. It is analogous to the phase transition in statistical mechanics, and the line of $k=2$ serves as the critical surface, which separates phase I (where Cycle 2 is global attractive) and phase II (where there is a positive but finite ratio between the two cycle fluxes) (Fig. \ref{fig1} (a)).

Similar behavior can be observed in the stochastic model approaching the deterministic counterpart with $\delta=0$, as $\alpha, \beta$ and $-\gamma$ are sufficiently large. Setting $\alpha=k\beta$ and $\gamma=-l\beta$, we can show that as long as $k+l<2$, the Cycle 2 is globally attractive; while as $k+l>2$, the probability of the fixed point is about $\frac{2}{3}$(Fig. \ref{fig1} (b)), given $\beta=50$. Analogous to phase transition, the two phases are separated by the critical surface $k+l=2$.

The phase transition in such a Boolean network model can be rigorously and quantitatively explained by the mathematical theory of exponentially perturbed Markov chains \cite{Chen1994,SM}. Within this type of stochastic models, we can properly define the concepts of nonequilibrium ``activation energy barrier" along each transiting trajectory from one attractor, either fixed point or limit cycle, to the other one. Then the theorem in \cite{Chen1994} tells that the mean transiting time $\tau_{ij}$ from the $i$-th attractor to the $j$-th one, satisfies

\begin{equation}
  \lim_{\beta\rightarrow\infty}\frac{1}{\beta}\log\langle\tau_{ij}\rangle=\Delta\Phi_{ij},
\end{equation}
where $\Delta\Phi_{ij}$ is the minimum activation energy barriers along all the transiting trajectories. See \cite{SM} for the expression of $\Delta\Phi_{ij}$.

Fig. \ref{fig1}(c), (d) present the minimum activation energy barrier difference $\Delta\Delta\Phi_{21}=\Delta\Phi_{21}-\Delta\Phi_{12}$ with $\delta=1$ and $0$ respectively. Comparing these figures with  Fig. \ref{fig1}(a), (b), we can easily conclude the direct relationship between minimum activation energy barrier difference and the limiting cycle fluxes: $\Delta\Phi_{12}<\Delta\Phi_{21}$ means the dominance of Cycle 2 (Phase I), implying $\frac{\tau_{21}}{\tau_{12}}$ tends to positive infinity with an exponentially increasing rate as $\beta\rightarrow\infty$; while $\Delta\Phi_{12}=\Delta\Phi_{21}$ indicates a nontrivial probability ratio between the two cycles (Phase II). In this example, we do not consider the case $\Delta\Phi_{12}>\Delta\Phi_{21}$, but it is definitely possible with a different $T$ matrix (See below).

Corresponding to the case $\delta=1$, the minimum activation energy barriers $\Delta\Phi_{12}=\min(k,2)$ and $\Delta\Phi_{21}=2$, even independent of $l$. Hence phase I in which the Cycle 2 is globally attractive, occurs if and only if $k<2$. On the other hand, corresponding to the case with $\delta=0$, the minimum energy barriers $\Delta\Phi_{12}=\min(k+l,2)$ and $\Delta\Phi_{21}=2$, therefore phase I occurs in the region of $k+l<2$. The theory clearly explains the origin of the two phases in such stochastic Boolean network models.

The theorem above cannot give the quantitative ratio of cycle fluxes in phase II, as the minimum activation energy barriers forward and backward are equal. However, according to the theory in \cite{Chen1994}, we have already known that as the noise is sufficiently low, the transiting events between the two attractors almost surely concentrate on the optimal transition paths, namely the transiting trajectories with the minimum activation energy barrier.

In such specific stochastic Boolean network models, we can prove something more: the probability of taking each optimal transition path is exactly the same, hence the ratio of mean transiting time satisfies \cite{SM}
\begin{equation}
  \lim_{\beta\rightarrow\infty}\frac{\langle\tau_{12}\rangle}
  {\langle\tau_{21}\rangle}=\dfrac{|A_1|\cdot n_{21}}{|A_2|\cdot n_{12}},\label{eq4}
\end{equation}
where $A_i(i=1,2)$ denotes the number of states in the $i$-th attractors of the deterministic models and $n_{12}(n_{21})$ represents the corresponding number of optimal transition paths.

Combining (\ref{eq4}) and the ergodic theorem, we deduce that the net fluxes of the two cycles $c_1$ and $c_2$ satisfy  \cite{SM}
\begin{equation}
\begin{cases}
|A_1|\cdot c_1+|A_2|\cdot c_2=1;\\
\dfrac{c_1}{c_2}=\dfrac{n_{21}}{n_{12}}.
\end{cases}
\end{equation}

The numbers of optimal transition paths and the corresponding net fluxes of the two cycles in Phase II are listed in Table \ref{tab_n}, which perfectly agree with the numerical values obtained from Fig. \ref{fig1} (a)(b).

\begin{table}[!hbp]
\begin{tabular}{|c|c|c|c|c|}
\hline
&\multicolumn{2}{|c|}{$\delta=1$($|A_1|=10$,$|A_2|=4$)} & \multicolumn{2}{|c|}{$\delta=0$($|A_1|=1$,$|A_2|=4$)} \\
\hline
&$k>2$& $k=2$& $k+l>2$ & $k+l=2$ \\
\hline
$n_{12}$& 10& 12 & 1& 3\\
\hline
$n_{21}$& 8& 8 & 8& 8\\
\hline
$c_1$&1/15&1/16&2/3&2/5\\
\hline
$c_2$&1/12&3/32&1/12&3/20\\
\hline
\end{tabular}
\caption{Optimal transition paths in Phase II.}
\label{tab_n}
\end{table}

\begin{figure}[bht]
\centerline{\includegraphics[width=8.5cm]{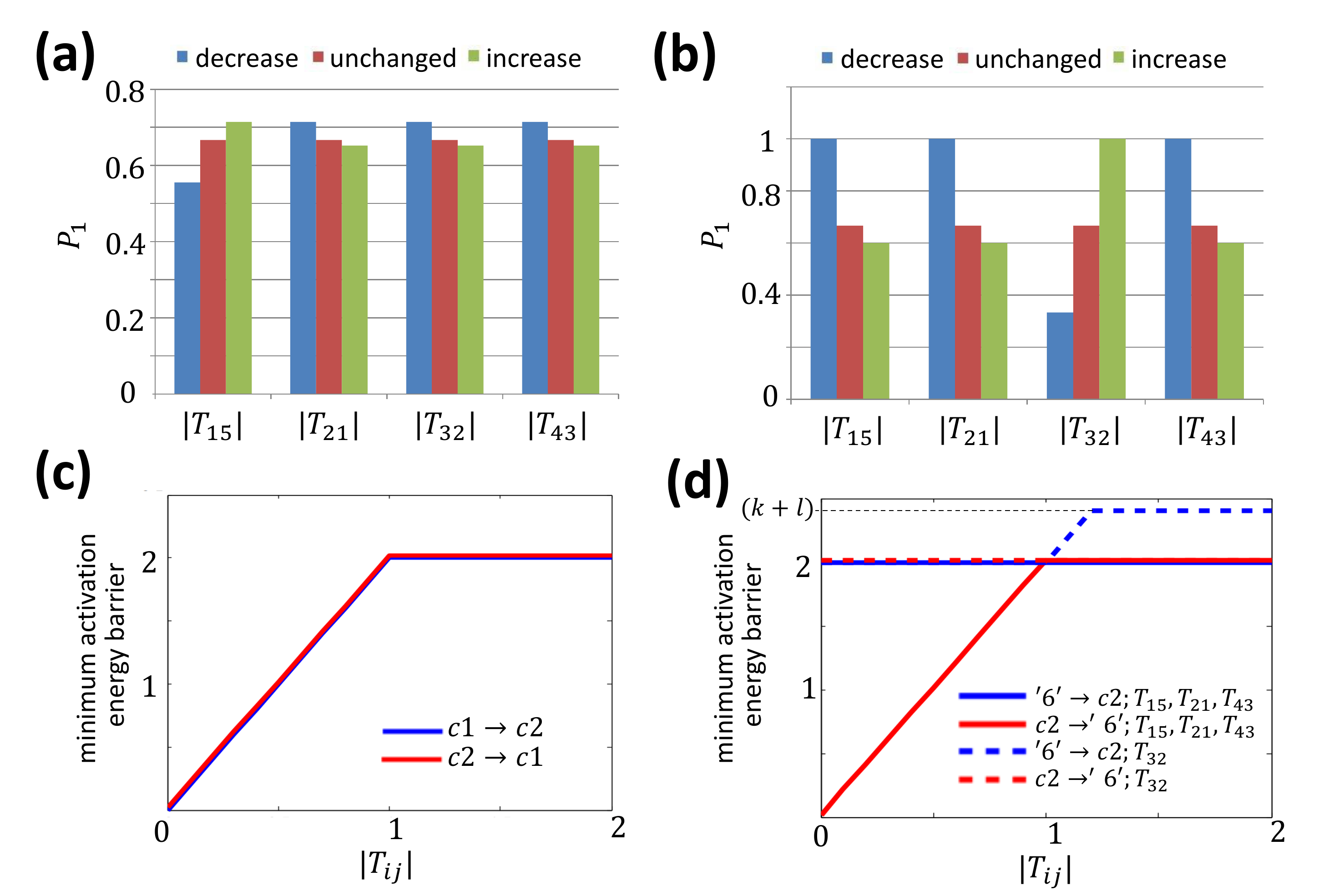}}
\caption{Probability of cycle 1 or node ``6" (i.e. $P_1=|A_1|\times c_1$) as well as the minimum activation energy barriers under slight changes in nonzero components of $T$. (a)(b) $P_1$ as each nonzero component is increased (blue) or decreased (green) in the case $\delta=1$, $k>2$ (a), and $\delta=0$, $k+l>2$ (b). (c) Minimum activation energy barrier from cycle 1 to cycle 2 (blue) and that from cycle 2 to cycle 1 (red) when a nonzero component changes in $[0,2]$, in the case $\delta=1$, $k>2$. They completely overlap with each other. (d) Minimum activation energy barrier from node ``6" to cycle 2 (blue) and that from cycle 2 to node ``6" (red) when each nonzero component of $T$ changes in $[0,2]$, in the case $\delta=0$, $k+l>2$. Note that the blue solid line entirely overlaps with the red dashed line. In plotting this figure we take $k+l=2.4$.}
\label{fig4}
\end{figure}

Our theory can also determine how the transition time and the number of optimal transition paths between parallel pathways depend on each interaction's strength, as well as can identify those possibly more crucial interactions in the biological networks. Because there are multiple optimal transition paths between the parallel pathways,  small changes of most single nonzero components in the interaction matrix $T$ can not significantly change the behaviors of the network. For example, if we consider the probability of the normal steady state and normal pathway of cells before or after being perturbed, which is crucial for the maintaining of the cellular functions, none of the interactions in $T$ dramatically affect the qualitative behaviors of the network in the global attractive region (Phase I) for both cases of $\delta=0$ and $1$ as well as the non-global attractive region (Phase II) for the case $\delta=1$,  as long as the strength of the interaction only varies within a small fraction (Fig. \ref{fig4} (a))\cite{SM}. However, in the non-global attractive region (Phase II) when $\delta=0$, a slightly varying of single nonzero component in $T$ can result in a qualitative change of the relative stability: from Phase II transiting to a new phase (Phase III) in which the normal steady state is globally attractive, which is shown in Fig. \ref{fig4} (b). It indicates that decreasing the interaction strengths from  p53 to Siah-1, Mdm-2 to p53 and $\beta$-catenin to p19/14ARF or increasing the interaction strength from Siah-1 to $\beta$-catenin can possibly make the normal cell more robustly against stochastic perturbations. It results from the fact that some interactions only affect the transition from one attractor to another but not on the opposite direction. Fig. \ref{fig4} (c), (d) show the value of minimum activation energy barriers along both directions with respect to several nonzero elements of $T$ under the conditions of Fig. \ref{fig4} (a) and (b) respectively. Both of the forward and backward minimum activation energy barriers increase with each interaction strength but possibly having different slopes. Also, even the topology of the networks can be significantly changed due to the modification of certain interaction strengths, such as the $T_{51}$ and $T_{54}$(Fig. \ref{fig0} (b)). See \cite{SM} for more details.

As a conclusion, parallel pathways often exist in biological systems, and their relative stability can be of great importance for implementing the correct cellular functions. These properties can not be understood and analyzed through simply observing the topology of the network or even studying deterministic models. We demonstrate here through stochastic Boolean network models of Siah-1/beta-catenin/p14/19 ARF loop that, there can possibly be various phases with qualitatively different behaviors of relative stabilities in such a nonequilibrium system. It is closely related to the nonequilibrium activation energy barriers along the transiting trajectories between these pathways, as well as the number of optimal transition paths with the minimum activation energy barrier. It can also help to identify which interactions in the network can dramatically affect the behaviors of the network, either making it better or worse. Our theory indicates that stochastic models, combined with the idea of nonequilibrium statistical mechanics, can help to address the stochastic robustness and relative stability in biological networks, at least semi-quantitatively. More implication of our theory remains to be further elucidated, by performing  suitable experiments as well as by modeling of the other important biological systems.

We thank Das Biswajit, Xiao Jin and Chen Jia for carefully reading the manuscript. This research is financially supported by NSFC (Nos. 10901040, 21373021) and the Foundation for the Author of National Excellent Doctoral Dissertation of China (No. 201119), to which we would like to express our genuine gratitude.

\end{document}